\documentclass[onecolumn,showpacs,preprintnumbers,showkeys,superscriptaddress]{revtex4}
\usepackage{graphicx}
\usepackage{dcolumn}
\usepackage{bm}

\def\eqref#1{Eq.~(\ref{eq:#1})}

\begin{document}

\title{Lowest Eigenvalues of Random Hamiltonians}
\author{J. J. Shen}
\affiliation{Department of Physics,  Shanghai Jiao Tong
University, Shanghai 200240, China}
\author{Y. M. Zhao}     \email{ymzhao@sjtu.edu.cn}
\affiliation{Department of Physics,  Shanghai Jiao Tong University,
Shanghai 200240, China}  \affiliation{Center of Theoretical Nuclear
Physics, National Laboratory of Heavy Ion Accelerator, Lanzhou
730000, China} \affiliation{Nishina Center, Institute of Physical
Chemical Research (RIKEN), Hirosawa 2-1, Wako-shi,  Saitama
351-0198, Japan}\affiliation{CCAST, World Laboratory, P.O. Box 8730,
Beijing 100080, China}
\author{A. Arima}
\affiliation{Science Museum, Japan Science Foundation, 2-1
Kitanomaru-koen, Chiyoda-ku, Tokyo 102-0091, Japan}
\author{N. Yoshinaga}
\affiliation{Department of Physics, Saitama University, Saitama
338-8570, Japan}

\date{\today}

\begin{abstract}
In this paper we present results of  the lowest eigenvalues of
random Hamiltonians  for both fermion  and boson systems. We show
that an empirical formula of evaluating the lowest eigenvalues of
random Hamiltonians in terms of energy centroids and  widths of
eigenvalues are applicable to many different systems (except for $d$
boson systems). We improve the accuracy of the formula by adding
moments higher than two. We suggest another new formula to evaluate
the lowest eigenvalues for random matrices with large dimensions
(20-5000). These empirical formulas are shown to be applicable not
only to the evaluation of the lowest energy but also to the
evaluation of excited energies of systems under random two-body
interactions.
\end{abstract}


\vspace{0.4in}

\maketitle

\newpage

\section{Introduction}

\indent The  $I^дл = 0^+$ ground state dominance for random
Hamiltonians was discovered by Johnson et al. in 1998
\cite{Johnson}.  Many efforts have been devoted to understand this
problem since then.  See Refs. \cite{review1,Zele} and references
therein for details.

Recently, Papenbrock and Weidenmueller considered fluctuations of
and correlations between the $J$-dependent spectral widths
\cite{Papenbrock1,Papenbrock2}. By using such correlations they were
able to approximately evaluate spin $I$ ground state probabilities
for six nucleons in a single-$j$ ($j=19/2$) shell under a two-body
random ensemble (TBRE). Along this line, a very simple formula of
the lowest energy of spin $I$ states based on energy cenroid,
spectral width and dimension of spin $I$ states, was presented in
Ref.\cite{Yoshinaga} by  the present authors. Our formula (see Eq.
(12) of Ref. \cite{Yoshinaga}) was shown to be applicable to
evaluate statistically the lowest energy of spin $I$ states, and
proved to be very good in predicting spin $I$ ground state
probabilities calculated by using random two-body interactions. We
should note that the idea of evaluating the lowest eigenvalue based
on energy centroids and spectral widths was suggested by Ratcliff
\cite{Ratcliff}, Vary {\it et al.} \cite{Vary}, and by Zuker and his
collaborators \cite{Zuker}. However, predicted results by  formulas
in Refs. \cite{Papenbrock1,Papenbrock2,Ratcliff,Vary,Zuker} are less
accurate than the formula in Ref. \cite{Yoshinaga}.

Evaluation of the lowest eigenvalue is not only useful in studying
regular structure of atomic nuclei in the presence of random
interactions, but also a very common practice in many other fields.
It is therefore the purpose of this article to revisit empirical
formulas of evaluating the lowest eigenvalues  under random
Hamiltonians. In this paper we shall also suggest other empirical
formulas of evaluating the lowest eigenvalues of random Hamiltonians
and random matrices.

This paper is organized as follows. In Sec. II we review  results of
evaluating the lowest eigenvalues by using energy centroids and
width. It is found that the results of single-$j$ shell and
$sd$-boson systems follow the formula of Ref. \cite{Yoshinaga}
(except $d$ bosons). In Sec. II we also improve such evaluations by
adding higher moments for various systems. In Sec. III we
investigate other four lowest eigenvalues. In Sec. IV we concentrate
on discussions of $d$ bosons. The lowest eigenvalues of random
matrices are discussed in the Appendix.

\section{empirical formula of Ref. \cite{Yoshinaga}}

In this paper we take the same notations as our earlier work
\cite{Yoshinaga}. For fermions in a single-$j$ shell
\begin{equation}
\widehat{H}=\sum_{J=0,\rm even}^{2j-1}\sqrt{2J+1}G_J \left[ A^{J
\dagger } \times \widetilde{A}^J \right]^0,
\end{equation}
with $A^{J \dagger }=\frac{1}{\sqrt{2}}[a_j^\dagger a_j^\dagger]^J$
and $\widetilde{A}^J=-\frac{1}{\sqrt{2}}[\widetilde{a}_j
\widetilde{a}_j]^J$. Two-body matrix elements $G_J$'s are assumed to
follow the TBRE, i.e., they are a set of random numbers with a
distribution function
\begin{equation}
\rho(G_J)=\frac{1}{\sqrt{2\pi}}{\rm exp}(-G_{\it J}^2/2),\, \it
J=\rm 0,2,...,2\it j-\rm 1.
\end{equation}
Matrix elements of $\widehat{H}$ for spin-$I$ states can be
expressed in terms of coefficients of fractional parentage (cfp's):
\begin{equation}
H_{I\beta\gamma}=\langle j^nI\beta|\widehat{H}|j^n
I\gamma\rangle=\sum_J \alpha^J_{I\beta\gamma}G_J.
\end{equation}
One can obtain eigenenergies of spin $I$ states  by diagonalizing
$H_{I\beta\gamma}$. Such definition can be easily generalized to
many-$j$ shells without confusion.

In Ref. \cite{Yoshinaga}, a very simple formula of evaluating the
lowest energy of $H_I$ was suggested as follows,
\begin{equation}
E_I^{(\rm min)}=\bar{E}_I-\Phi(d_I)\sigma_I , \label{coef}
\end{equation}
where $\Phi(d_I) =\sqrt{a\ln{d_I}+b}$, $\sigma_I$ is the square root
of the second order moment of eigenvalues of all spin $I$ states,
$d_I$ is dimension of spin $I$ states, and $a$ and $b$ were
determined empirically to be $0.99$ and $0.36$, respectively. This
formula was found to hold statistically.

Let us first investigate fermion systems. We take four fermions in a
$j=15/2$ shell, four fermions in a $j=21/2$ shell, four fermions in
a $j=31/2$ shell, five fermions in a $j=19/2$ shell, six fermions in
a $j=17/2$ shell, and four fermions in a two-$j$ ($j=7/2,5/2$)
shell. The procedure is the same as in Ref. \cite{Yoshinaga}; first,
$\Phi(d_I)=[\bar{E}_I-E^{(\rm min)}_I]/\sigma_I$ are calculated for
1000 runs of the TBRE, and second, we calculate the average of
$\Phi(d_I)$ for these 1000 runs. We plot  $\left( \Phi(d_I)
\right)^2$ as a  function of ln$d_I$ in Fig. 1.  One can see that
$\left( \Phi(d_I) \right)^2$ is close to linear correlation with
$\ln(d_I)$ for all these examples. Parameters $a$ and $b$  obtained
by this procedure in each example are shown in Table I. They are
close to those obtained in Ref. \cite{Yoshinaga} ($a=0.99, b=0.36$).
The results of $sd$ bosons are similar to those of fermions in a
single-$j$ shell or two-$j$ shells, as shown in Fig. 1(b) and Table
II. We also investigate results of a more complicated system: three
valence protons and three valence neutrons in the
$(2s_{1/2},1d_{3/2},0i_{11/2})$ shell, as shown in Fig. 2.

\begin{center}
\includegraphics[width = 6.0in]{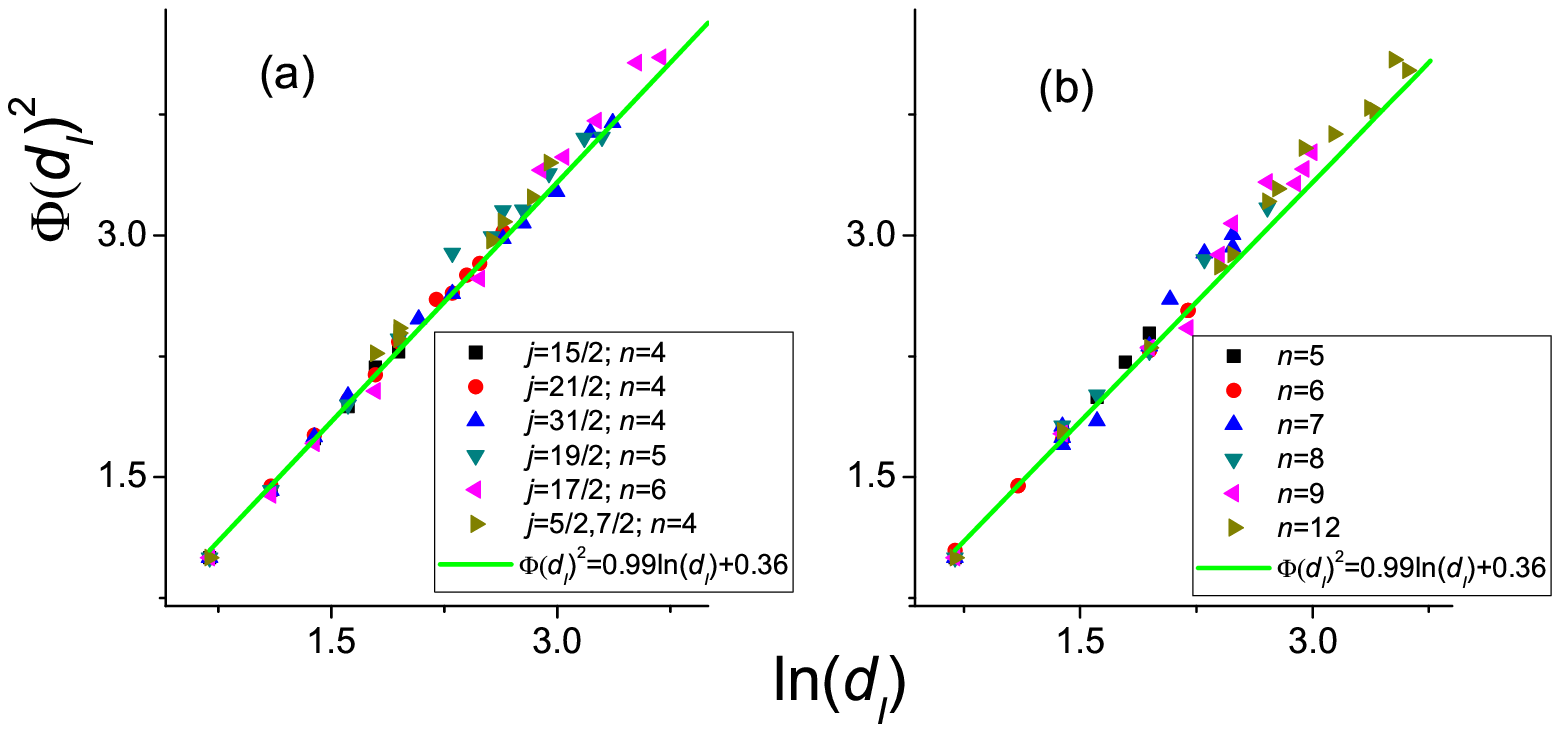}
\end{center}

FIG. 1.(Color online) Factor $(\Phi(d_I))^2$ versus $\ln{(d_I)}$
determined numerically for fermion and sd-boson systems. (a)~
fermion systems; (b) ~ $sd$-boson systems. The dimension in these
cases is 2-40.

\vspace{0.4in}

{TABLE I. $a$ and $b$ values for fermions in a single-$j$ shell and
two-$j$ shells corresponding to (a)-(b) of Fig. 1(a).  }

\vspace{0.1in}

\begin{small}
\begin{tabular}{cccccccc}
\hline  \hline ($2j,n)$ & (15,4) & (21,4) &(31,4)& (19,5)& (17,6)&
($2j_1=5,2j_2=7,n=4$) & Ref. \cite{Yoshinaga}
\\ \hline
$a$ & $1.04 \pm 0.03$ & $1.02 \pm 0.01$& $1.00\pm 0.02$  &
$1.04\pm0.03$ & $1.07\pm0.02$  & $1.06\pm 0.02$ & 0.99
\\
$b$ & $0.28\pm0.04$ &  $0.32\pm0.03$& $0.34\pm 0.04$  &
$0.32\pm0.08$& $0.20\pm0.06$  & $0.30\pm 0.04$ & 0.36 \\ \hline
\hline
\end{tabular}
\end{small}

\vspace{0.4in}

{TABLE II. $a$ and $b$ values of $sd$-boson systems with $n=5$, $6$,
$7$, $8$, $9$, and $12$, respectively. See Fig. 1(b). }

\vspace{0.1in}

\begin{small}
\begin{tabular}{ccccccc}
\hline  \hline $n$ &5 &  6 & 7 & 8 & 9 & 12
\\ \hline
$a$ & $1.10\pm0.02$ & $1.09\pm0.03$& $1.13\pm0.05$ & $1.08\pm0.04$&
$1.10\pm0.05$& $1.05\pm0.02$
\\
$b$ & $0.24\pm0.03$ &  $0.25\pm0.05$ & $0.17\pm 0.08$&
$0.26\pm0.08$& $0.22\pm0.11$& $0.30\pm 0.07$
\\
\hline  \hline
\end{tabular}
\end{small}

\begin{center}
\includegraphics[width = 4.0in]{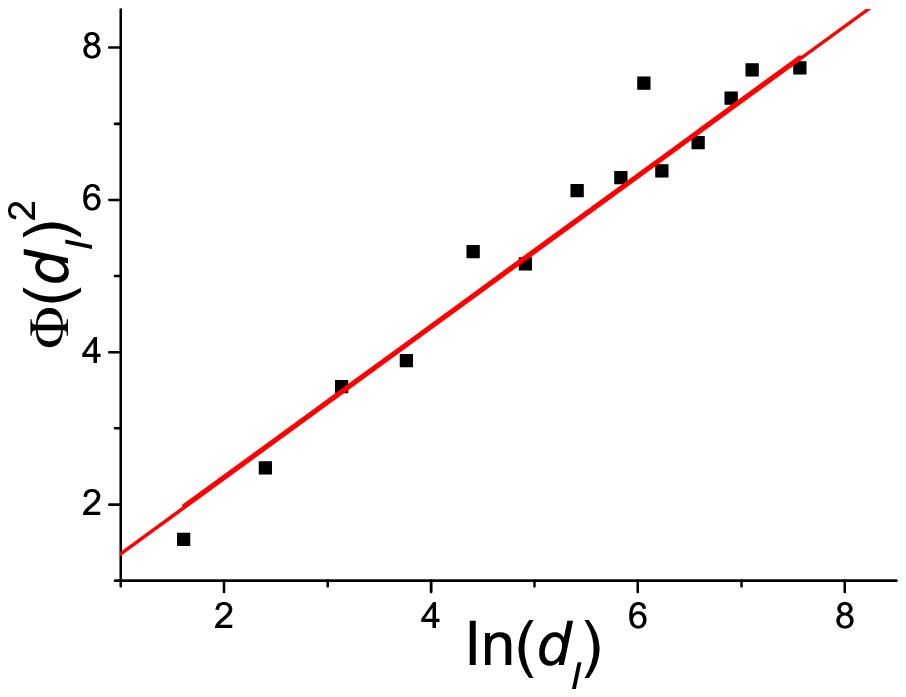}
\end{center}
FIG. 2.(Color online)  Factor $(\Phi(d_I))^2$ versus $\ln{(d_I)}$
for three valence protons ($N_p=3$) and three valence neutrons
($N_n=3$) in the shell $(2s_{1/2},1d_{3/2},0i_{11/2})$. The line is
plotted by using $\Phi(d_I)^2 =  {a\ln{d_I}+b}$.

\vspace{0.2in}

Eq. (\ref{coef}) was shown to be well applicable for  evaluating
spin $I$ g.s. probability  (see Ref. \cite{Yoshinaga} for details).
However, it is not  good enough for evaluation of the ground state
energy in a reasonable precision. It is therefore desirable to
improve the formula of Ref. \cite{Yoshinaga}.

Towards this goal, we consider higher orders of moment of the
eigenenergies to compensate the deviation from Gaussian distribution
for eigenvalues. In this paper we consider the cubic root of the
third order of central moment, denoted by $\sigma_3$:
\begin{equation}
\sigma_3 = \left(\frac{1}{d_I} \sum_{i=1}^{d_I}
(\sum_{k=1}^{d_I}H_{ki} (\sum_{j=1}^{d_I}H_{ij}H_{jk}))\\
-3\frac{1}{d_I^2} (\sum_{i=1}^{d_I}(\sum_{j=1}^{d_I}H_{ij}^2)
\sum_{i=1}^{d_I} H_{ii}+2\frac{1}{d_I^3}
(\sum_{i=1}^{d_I}H_{ii})^3\right)^{1/3} ~.
\end{equation}
Let us assume
\begin{equation}
E_I^{(\rm min)}=\bar{E}_I-C_2\sigma_2+C_3\sigma_3,~ \label{pred}
\end{equation}
where $\sigma_2=\sigma_I$ in Eq. (\ref{coef}). According to our
numerical experiments (see Table III), the disagreement between
predicted result of $E_I^{(\rm min)}$ by using Eq. (\ref{pred}) and
that by diagonalizing  $H_I$ can be reduced to about $1/2$ on
average, in comparison with our earlier formula Eq.(\ref{coef}).
Unfortunately, we are not able to obtain a simple formula for $C_3$
. The values of $C_2$ such obtained are very close to the value of
$\Phi(d_I)$.

 \vspace{0.15in}

 {TABLE III. Relative
deviation of Eqs. (\ref{coef}) and (\ref{pred}) for four fermions in
a single-$j$ shell with $j=31/2$ and $n=4$. Error ($\epsilon$) is
the TBRE average calculated by the value of $\left|\frac{E_I^{\rm
pred}-E_I^{\rm exact}}{E_I^{\rm exact}}\right|$. Error A
($\epsilon_A$) is obtained by  Eq. (\ref{pred}), while   Error B
($\epsilon_B$) is obtained by  Eq. (\ref{coef})}.

\vspace{0.2in}

\begin{small}
\begin{tabular}{cccccccccc} \hline  \hline
$d$ & 3  & 4  & 5 & 8 & 10 & 14 & 20 & 25 & 29  \\
$I$ & 50 & 3  & 0 & 39& 2  &  4 & 13 & 12 & 20 \\ \hline
$\epsilon_A$ & 0.40 & 0.46 & 0.18 & 0.28 & 0.33 & 0.30 & 0.13 & 0.15
& 0.17 \\
$\epsilon_B$ & 0.31 & 0.17 & 0.098 & 0.18 & 0.15 & 0.12 & 0.091 &
0.089 & 0.10
\\    \hline  \hline
\end{tabular}
\end{small}

\vspace{0.3in}
\section{Other lowest energies }

The success of evaluating the lowest energies encourages us to go
further. Here we study eigenvalues of the first to the fourth
excited states. Typical results are shown in Fig. 3, where we take
the same form of  Eq. (\ref{coef}). One can confirm here that Eq.
(4) is also applicable to evaluate the excited energies for both
fermions and $sd$ boson systems while the values of $a$ and $b$ are
different from the ground state. The results are summarized in Fig.
4 and Table IV.

.

\begin{center}
\includegraphics[width = 6.0in]{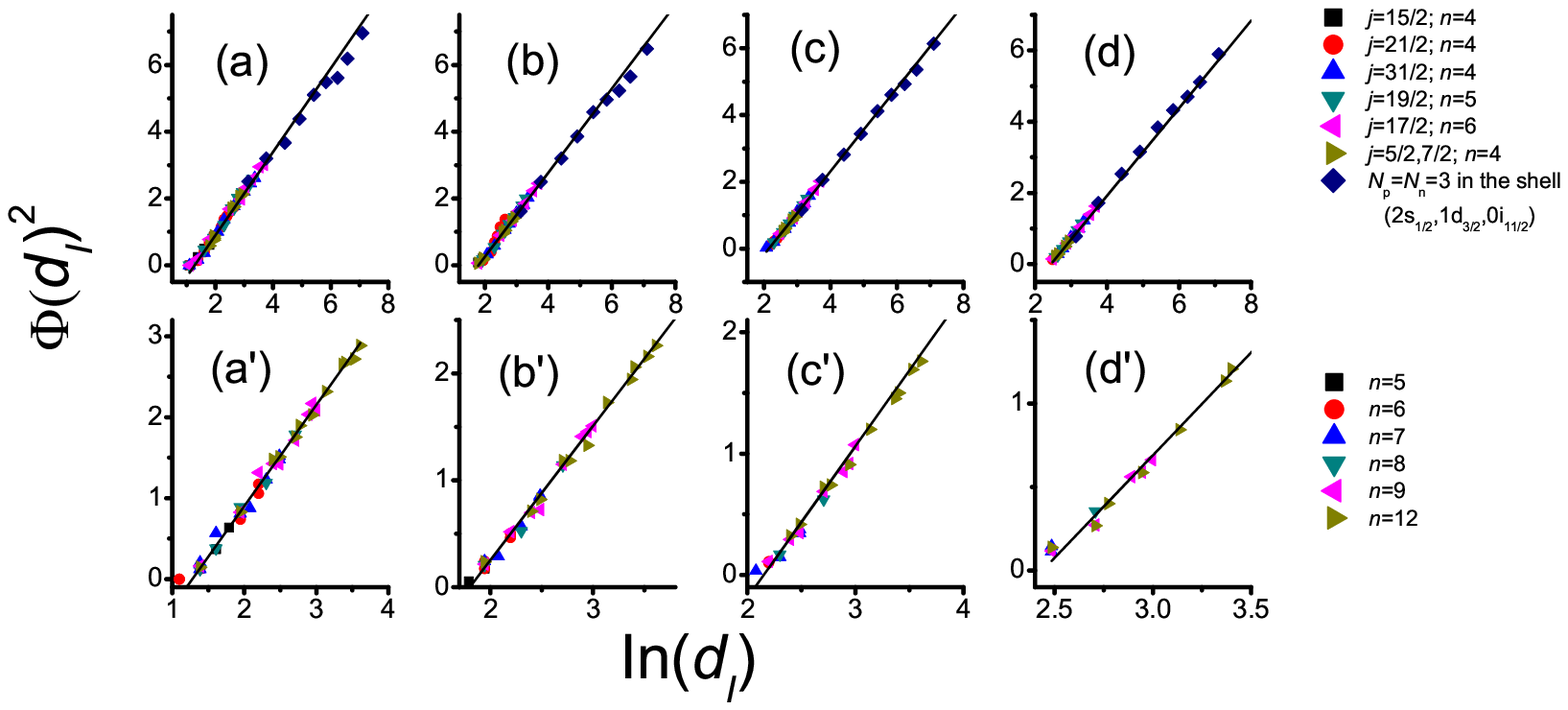}
\end{center}

FIG. 3. (Color online) $[\Phi(d_I)]^2$ versus $\ln(d)$. (a)~ the
first excited states for fermions. (a$'$)~ the first excited states
for $sd$ bosons. (b)~ the second excited states for fermions.
(b$'$)~ the second excited states for $sd$ bosons. (c)~ the third
excited states for fermions. (c$'$)~ the third excited states for
$sd$ bosons. (d)~ the fourth excited states for fermions. (d$'$)~
the fourth excited states for $sd$ bosons.
\\
\\
\\

\begin{center}
\includegraphics[width = 4.0in]{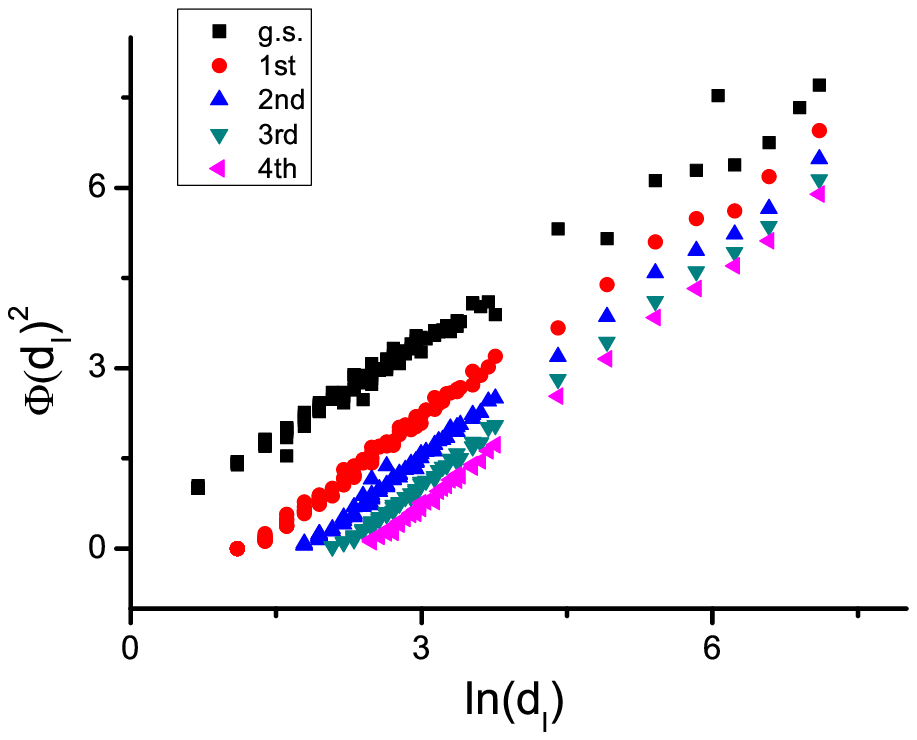}
\end{center}

FIG. 4. Comparison of average $\Phi(d_I)^2$  versus $\ln(d)$. In the
figure, ``g.s." means the ground state, ``1st" means the first
excited state, ``2nd" means the second excited state, ``3rd" means
the third excited state, and ``4th" means the fourth excited state.

 \vspace{0.4in}

It is worthy to note that our empirical formulas  are based on
calculations with finite  dimensions (dimension is less than
$10^4$). According to the linear correlation between $\left
(\Phi(d_I) \right)^2$ and ln$(d_I)$ shown in Fig. 4, the straight
line corresponding to  ground states and that corresponding to
excited states (e.g., the 1st excited states) seem to cut across
each other. Such an intersection does not occur, because the g.s.
energy  always corresponds a larger $\Phi$ value. In other words,
the linear correlation between  $\left (\Phi(d_I) \right)^2$ and
ln$(d_I)$ is valid when $d_I$ is smaller than $10^4$. For larger
$d_I$, the results should be further investigated.

 \vspace{0.4in}

{TABLE IV. $a$ and $b$ values for fermion systems and $sd$-boson
systems with eigenvalues of the ground state and the first to the
fourth excited states corresponding to Fig. 4, respectively. }

 \vspace{0.2in}

\begin{small}
\begin{tabular}{cccccc}
\hline  \hline State  & g.s. &  1st   & 2nd  & 3rd   & 4th
\\ \hline
$a$ & $1.04\pm0.01$ & $1.18\pm0.01$& $1.21\pm0.01$ & $1.23\pm0.01$&
$1.25\pm0.01$
\\
$b$ & $0.30\pm0.03$ &  $-1.42\pm0.02$ & $-2.14\pm0.03$&
$-2.64\pm0.02$& $-3.03\pm0.02$
\\
\hline  \hline
\end{tabular}
\end{small}

\section{ d-boson systems}

In this Section we study very simple systems,  $d$ bosons, for which
eigen-energies are linear combinations of two-body matrix elements.
The two-body Hamiltonian of a $d$-boson system is given by
\begin{eqnarray}
&& H_d= H_0 + \sum_l\frac{1}{2}\sqrt{2l+1}c_l[(d^\dagger\times
d^\dagger)^l \times (\widetilde{d}\times\widetilde{d})^l]^0~.
\end{eqnarray}
 From Eqs.(2.79) and (2.82) of Ref. \cite{Arima}, we have
\begin{eqnarray}
  E=  && E_0+ \frac{1}{14} (4c_2+3c_4)   n_d(n_d-1)+
\frac{1}{70} (7c_0-10c_2+3c_4) \left[n_d (n_d+3) - \upsilon
(\upsilon+3)
\right] \nonumber  \\
&+& \frac{1}{14} (-c_2 + c_4)\left[I (I+1)-6 n_d \right]~.
\label{d-boson}
\end{eqnarray}

For  states with given $I$, the state with  $\upsilon$ being maximum
value or minimum (depending on the sign of $\beta'=\frac{1}{70}
(7c_0-10c_2+3c_4)$) is the lowest among all eigenvalues. For given
$n_d$ and $I$,
\begin{eqnarray}
E_{I\upsilon}=E({n_d, I})-\frac{1}{70} (7c_0- 10 c_2 + 3 c_4)
 \upsilon (\upsilon+3)~,
\end{eqnarray}
where  $E({n_d, I})$ is the same for all spin $I$ states.

Our results of $\Phi(d_I)$ versus ln($d_I$) are presented in Fig. 5.
There seems no simple relationship between $\Phi$ and ${\rm ln}d_I$.
Apparently,  there is systematic deviation from linear correlation
between $[\Phi(d_I)]^2$ and $\ln{(d)}$. The value of $\Phi(d_I)$
seems to saturate when the dimension goes to infinity.

\begin{center}
\includegraphics[width = 5.0in]{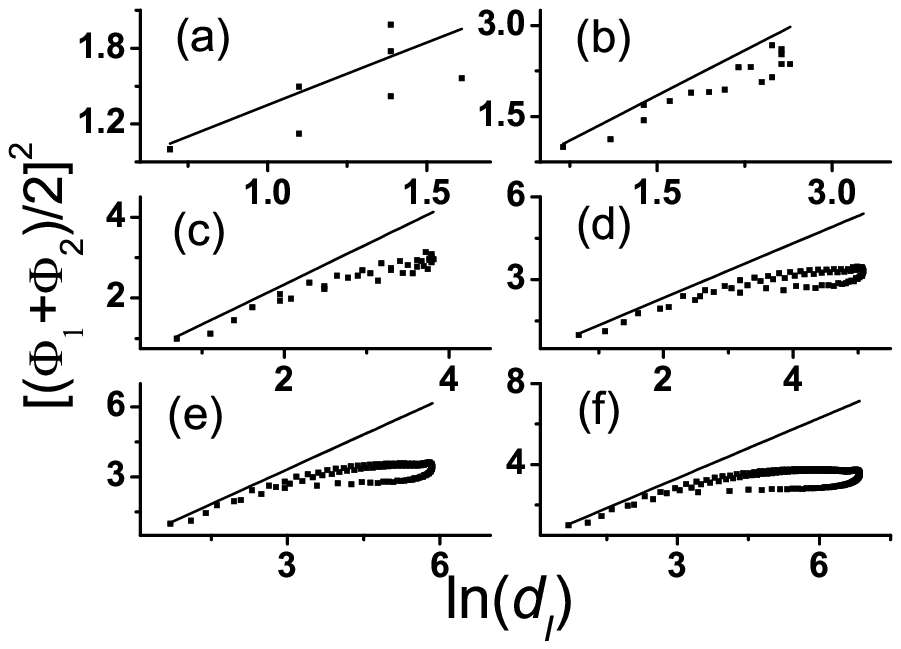}
\end{center}

Fig. 5.  Phenomenological factors $[\Phi(d_I)]^2$ versus $\ln{(d)}$
for $d$ bosons with different boson numbers. The line is plotted by
assuming  $[\Phi(d_I)]^2=0.99\ln{(d)}+0.36$.  (a) $n_d=9$, (b)
$n_d=18$, (c) $n_d=36$, (d) $n_d=72$,(e) $n_d=108$,(f) $n_d=180$.

 \vspace{0.3in}

The deviation from linear correlation between $\Phi(d_I)$ and
ln$d_I$ in Fig. 5 originates from the distribution of  eigenenergies
of $d$ bosons. The eigenvalues of other more complicated systems in
this paper exhibit a Gaussian distribution \cite{Mon}, while that of
$d$ bosons is close to a triangular ($I \geq n$) or  trapezoidal ($I
< n$) distribution, as shown  in Fig. 6.

\begin{center}
\includegraphics[width = 5.0in]{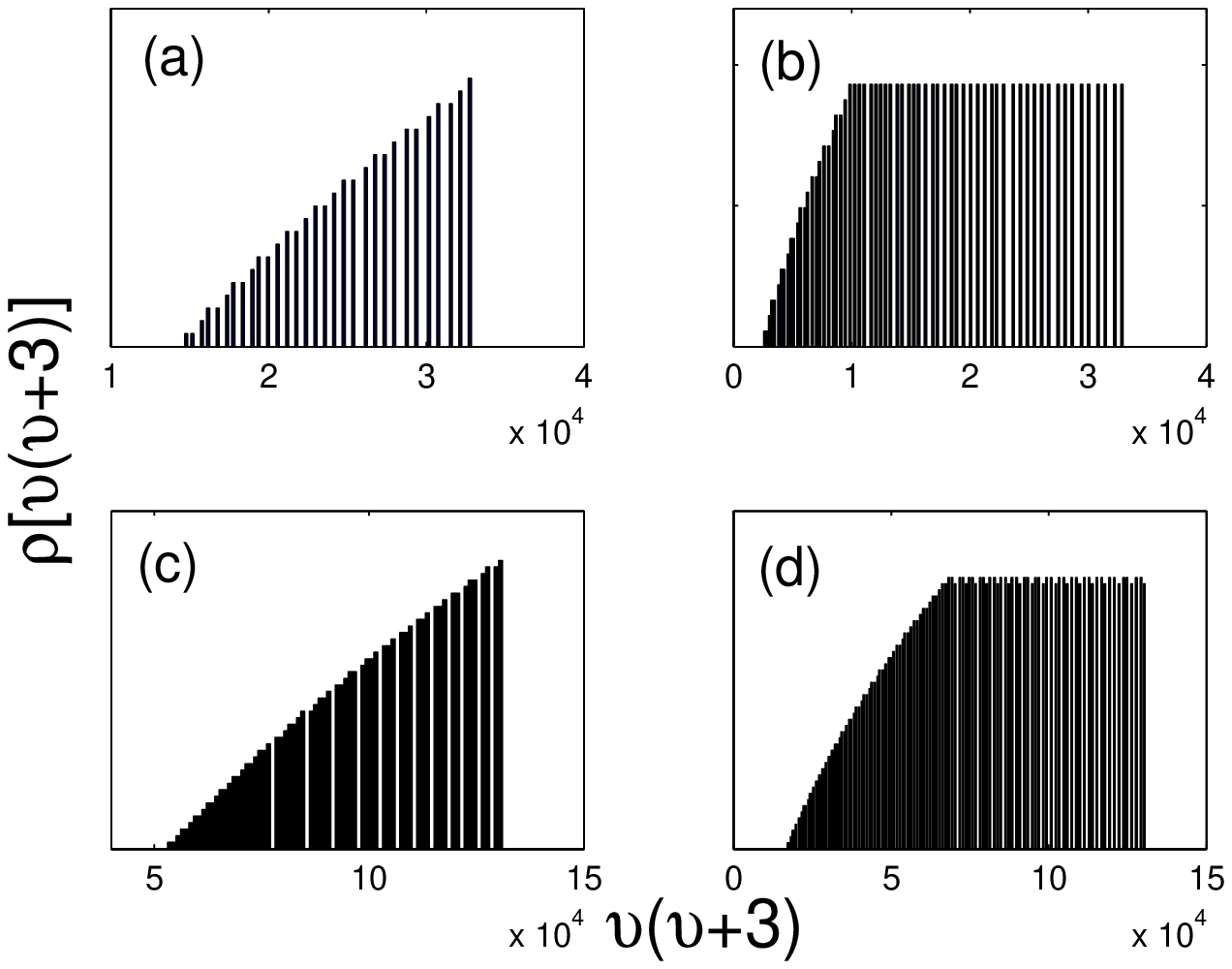}
\end{center}

Fig. 6. Relative distribution of eigenvalues of $d$ bosons.
  (a)~ $n=180, I=240$; (b)~ $n=180, I=100$;
(c)~ $n=360, I=460$;  (d)~ $n=360, I=260$.

\section{Summary and Discussion}

To summarize, in this paper we have studied lowest eigenvalues of
random Hamiltonians.  First, we demonstrate that our semi-empirical
formula suggested in Ref. \cite{Yoshinaga}, $E_I^{(\rm
min)}=\bar{E}_I-\Phi(d_I)\sigma_I$ with $(\Phi(d_I))^2=a{\rm ln}d+b$
works well for various examples ($sd$ bosons, fermions in a
single-$j$ or many-$j$ shells) with dimension ranging from 2 to 5000
(except for $d$ boson systems for which there are systematic
deviations). We also improve our formula by adding the third-order
central moment.

Second,  we investigate eigen-energies of the excited states
including the first, second, third and fourth lowest energies. We
find that the same formula with different parameters describes
statistically very well to eigen energies.

Third, we study $d$-boson  systems and discuss why there are
systematic deviation from our statistical formula of the lowest
eigenvalues.

In this paper we also study the lowest energies of random matrices
in Appendix. We see that there is systematic deviation from Wigner's
semi-circle prediction \cite{wigner} when the dimension is not very
large: when the dimension is less than 100, the lowest eigenvalue
can be evaluated statistically by the same formula of Ref.
\cite{Yoshinaga}. We suggest another statistical formula of lowest
eigenvalues of pure random matrices.

{\bf Acknowledgements:} We would like to thank the National Natural
Science Foundation of China for supporting this work under grants
10575070, 10675081. This work is also supported partly by the
Research Foundation Doctoral Program of Higher Education of China
under grant No. 20060248050, Scientific Research Foundation of
Ministry of Education in China for Returned Scholars,  the
NCET-07-0557, and by Chinese Major State Basic Research Developing
Program under Grant 2007CB815000.

\begin{center}
{\bf Appendix A}  The lowest eigenvalues of  random matrices
\end{center}

All  matrix elements in this Appendix are given by Gaussian
distributed random numbers. We change the dimension of matrices with
dimension ranging from 2 to 3000.  In Fig. 8(a), the dimension
changes from 2 to 15. In this case, we see a similar result as in
main text. There exists   linear correlation between $\Phi^2$ and
${\rm ln}d$ (see Fig. 8(a) ), with $a$ and $b$ are very close to
those in Sec. II. This means that Eq. (\ref{coef}) applies to
general cases (statistically) when dimension of matrix is not very
large.

When the dimension changes from 15 to 3000, we find that there is
systematic deviation from  linear correlation between $\Phi^2$ and
${\rm ln}d$. We empirically obtain
\begin{equation}
-\Phi(d)=a\ln{d}/d+b :\,a= 1.59\pm0.03,b=-2.00\pm0.003.
\label{random-mat}
\end{equation}
as shown in Fig. 8(b). We see that Eq. (\ref{random-mat}) is more
applicable than $[\Phi(d)]^2=a\ln{(d)}+b$  when dimension of matrix
becomes large. We note that the parameter $b$ in Eq.
(\ref{random-mat}) saturates at 2 in the large $d$ limit. This
saturation value is  the predicted value of Wigner's semi-circle
theorem.

\begin{center}
\includegraphics[width = 3.0in]{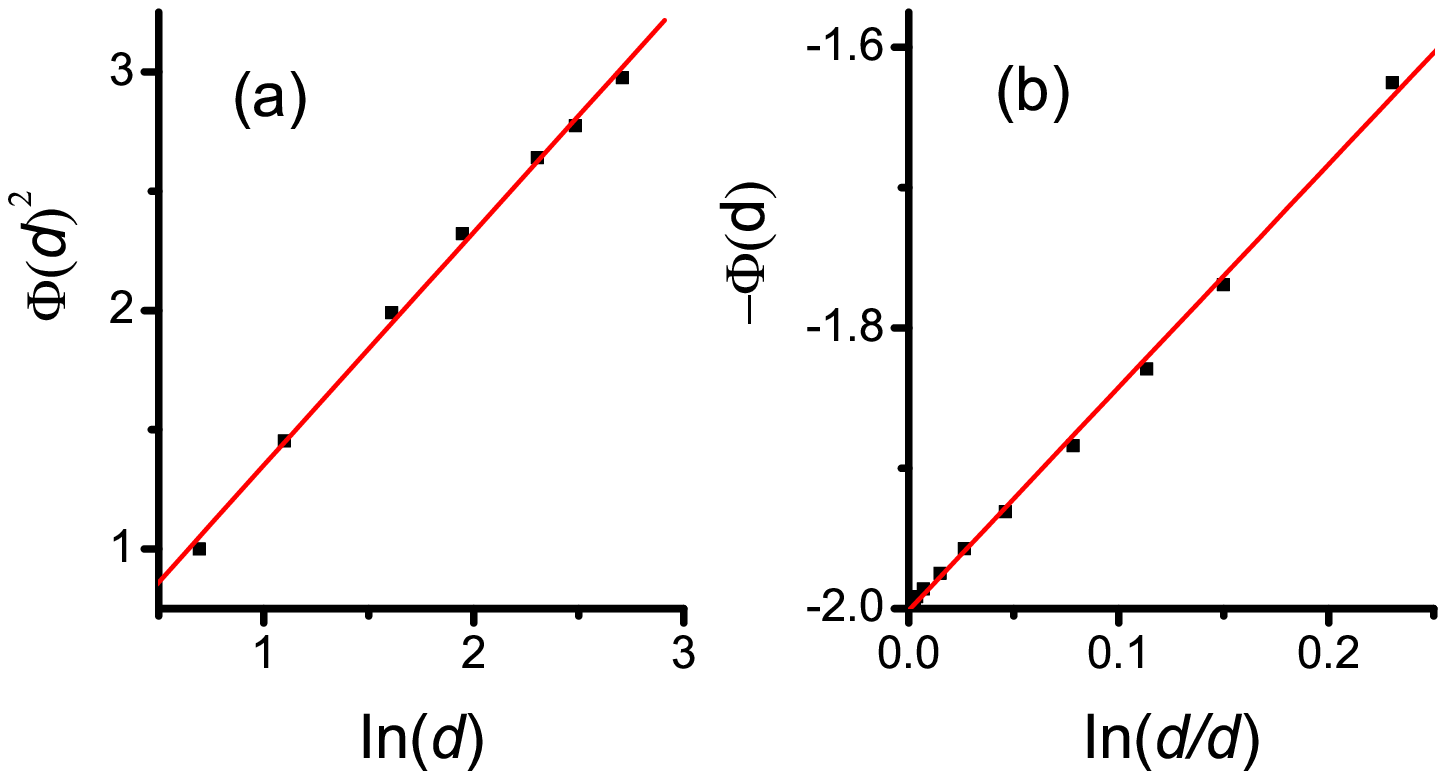}
\end{center}

Fig. 7. (Color online) Phenomenological factors $\Phi(d)$ versus
$\ln{(d)}$) for random matrices.  (a)~   matrices with smaller
dimension. Suppose $[\Phi]^2=a\ln{(d)}+b$. We obtain $a=0.97758\pm
0.02426, b=0.37201 \pm 0.04754$ by $\chi^2$ fitting. These values
are very close to those in Ref.\cite{Yoshinaga}; (b)~ matrices with
large dimension ($d=15-3000$).  One can see a nice linear
correlation between $-\Phi(d)$ and $\ln{d}/d$. We obtain that
$a=1.589\pm0.028$, $b=-2.001\pm0.003$ in Eq. (\ref{random-mat}).

\end{document}